\documentclass[prb,a4paper,10pt,twocolumn,showpacs,floatfix,superscriptaddress,amsmath,amsfonts,amssymb,preprintnumbers,longbibliography,citeautoscript]{revtex4-1}
\usepackage[T1]{fontenc}
\usepackage[utf8]{inputenc}
\usepackage{dcolumn,graphicx,color,booktabs,microtype,afterpage}
\graphicspath{{./}{figure/}}
\usepackage[charter,greekuppercase=italicized]{mathdesign}\usepackage{sidecap}

\renewcommand{\tablename}{Table}
\makeatletter\renewcommand{\fnum@table}[1]{\tablename~\thetable.}\makeatother
\newcount\hh \newcount\mm
\hh=\time \divide\hh by 60
\mm=\hh \multiply\mm by 60 \mm=-\mm
\advance\mm by \time
\def\now{\number\hh:\ifnum\mm<10{}0\fi\number\mm}

\usepackage[colorlinks,plainpages=false,linkcolor=blue,urlcolor=blue,citecolor=blue,pdfpagemode=UseNone,pdfstartview=FitBH]{hyperref}

\begin{document}

\makeatletter\renewcommand{\ps@plain}{%
\def\@evenhead{\hfill\itshape\rightmark}%
\def\@oddhead{\itshape\leftmark\hfill}%
\renewcommand{\@evenfoot}{\hfill\small{--~\thepage~--}\hfill}%
\renewcommand{\@oddfoot}{\hfill\small{--~\thepage~--}\hfill}%
}\makeatother\pagestyle{plain}

\preprint{\textit{Preprint: \today, \now.}}

\title{Microscopic\hspace{.3em}investigation\hspace{0.2em}of\hspace{0.2em}the\hspace{0.3em}weakly-correlated\hspace{0.2em}noncentrosymmetric\hspace{0.2em}superconductor\hspace{0.2em}SrAuSi$_3$}

\author{N.\ Barbero}
\affiliation{Laboratorium f\"ur Festk\"orperphysik, ETH Z\"urich, CH-8093 Zurich, Switzerland}

\author{P.\ K.\ Biswas}
\email{pabitra.biswas@stfc.ac.uk}
\affiliation{ISIS Pulsed Neutron and Muon Source, STFC Rutherford Appleton Laboratory, Harwell Campus, Didcot, Oxfordshire, OX11 0QX, United Kingdom}%

\author{M.\ Isobe}
\affiliation{Superconducting Properties Unit, National Institute for Materials Science, 1-1 Namiki, Tsukuba, Ibaraki 305-0044, Japan}%

\author{A.\ Amato}
\affiliation{Paul Scherrer Institut, CH-5232 Villigen PSI, Switzerland}%

\author{E.\ Morenzoni}
\affiliation{Paul Scherrer Institut, CH-5232 Villigen PSI, Switzerland}%

\author{A.\ D.\ Hillier}
\affiliation{ISIS Pulsed Neutron and Muon Source, STFC Rutherford Appleton Laboratory, Harwell Campus, Didcot, Oxfordshire, OX11 0QX, United Kingdom}%

\author{H.-R.\ Ott}%
\affiliation{Laboratorium f\"ur Festk\"orperphysik, ETH Z\"urich, CH-8093 Zurich, Switzerland}
\affiliation{Paul Scherrer Institut, CH-5232 Villigen PSI, Switzerland}

\author{J.\ Mesot}%
\affiliation{Laboratorium f\"ur Festk\"orperphysik, ETH Z\"urich, CH-8093 Zurich, Switzerland}
\affiliation{Paul Scherrer Institut, CH-5232 Villigen PSI, Switzerland}

\author{T.\ Shiroka}
\affiliation{Laboratorium f\"ur Festk\"orperphysik, ETH Z\"urich, CH-8093 Zurich, Switzerland}
\affiliation{Paul Scherrer Institut, CH-5232 Villigen PSI, Switzerland}%

\begin{abstract}
\noindent
SrAuSi$_3$ is a noncentrosymmetric superconductor (NCS) with $T_c$ = 1.54\,K, which to date has 
been studied only via macroscopic techniques. By combining nuclear magnetic resonance 
(NMR) and muon-spin rotation ($\mu$SR) measurements we investigate both the normal and 
the superconducting phase of SrAuSi$_3$ at a local level. 
In the normal phase, our data indicate a standard metallic behavior with weak electron 
correlations and a Korringa constant $S_\mathrm{exp} = 1.31 \times 10^{-5}$\,sK. The latter, 
twice the theoretical value, can be justified by  the Moriya theory of exchange enhancement. 
In the superconducting phase, the material exhibits conventional BCS-type superconductivity 
with a weak-coupling $s$-wave pairing, a gap value $\Delta(0)$ = 0.213(2)\,meV, and a 
magnetic penetration depth $\lambda(0)$ = 398(2)\,nm. 
The experimental proof of weak correlations in SrAuSi$_{3}$ implies that 
correlation effects can be decoupled from those of antisymmetric spin-orbit 
coupling (ASOC), thus enabling accurate band-structure calculations in the 
weakly-correlated NCSs.
\end{abstract}



\maketitle\enlargethispage{3pt}

\vspace{-5pt}\section{Introduction}\enlargethispage{8pt}
The absence of inversion symmetry in the crystal lattice
of certain materials introduces an antisymmetric spin-orbit coupling (ASOC) in the conduction-electron
ensemble. In case of superconductivity, this may result in mixed-parity phases,\cite{Sigrist2012}
which have increasingly attracted the attention of the scientific 
community, especially after the discovery of the first strongly-correlated NCS CePt$_3$Si.\cite{Bauer2004} 
With a $T_{\mathrm{N}}$ = 2.2\,K and a $T_c$ = 0.75\,K, it was argued 
that in CePt$_3$Si the magnetic fluctuations induce a non-standard 
pairing mechanism.

Since CePt$_3$Si, several new compounds belonging to different NCS 
classes have been discovered.\cite{Sigrist2012} In particular, analogous ternary compounds 
with \emph{weak} or \emph{negligible} electronic correlations have been 
found to exhibit conventional or mixed-pairing superconductivity (SC) and are actively 
under investigation.\cite{Smidman2017}
The study of NCSs with only weak electronic correlations is key to understanding 
the superconductivity in noncentrosymmetric compounds. Indeed, weak 
correlations make it possible to disentangle the ASOC, which reflects 
the lack of inversion symmetry, from the role of correlations.
Accurate band-structure calculations, focused exclusively 
on the ASOC-splitting of electronic bands, are therefore within reach.

SrAuSi$_3$, recently synthesized in the form of polycrystals 
by means of high-pressure techniques,\cite{Isobe2014} crystallizes in the BaNiSn$_3$-type 
NCS structure (tetragonal with space group $I4mm$, SG no.\ 107). The first  
magnetization and transport measurements determined the onset of superconductivity at $T_c$= 1.54\,K 
and an upper critical field $\mu_0H_{c2}(0)$ of 0.18\,T.
Subsequently, two theoretical first-principles calculations on SrAuSi$_3$ revealed lattice constants and
an electronic structure which are only in partial agreement with the experimental data.\cite{Shu2015,Shubis2015}
Further measurements of the specific heat and band calculations\cite{Isobe2016} suggested a nodeless gap, 
supporting the hypothesis that SrAuSi$_3$ is an $s$-wave BCS-type superconductor. Additional theoretical calculations\cite{Arslan2016} 
claimed an electron-phonon coupling parameter $\lambda$ = 0.47 and attributed the pairing mainly to the $p$ electrons of Si.
SrAuSi$_3$ may be compared with BaPtSi$_3$, an isostructural compound with very similar 
lattice parameters, Debye temperature, and electron-phonon coupling. Ref.~\onlinecite{Bauer2009} reports that 
BaPtSi$_3$ is a standard weak-coupling conventional superconductor with $2\Delta/k_{\mathrm{B}}T_c$ = 3.5. 

Below we report the first \emph{microscopic} study of SrAuSi$_3$ by employing nuclear 
magnetic resonance (NMR) and muon-spin rotation ($\mu$SR) measurements. We 
probed both its normal and superconducting phase and show that SrAuSi$_3$ 
is a weakly-correlated metal turning into a conventional superconductor.

Sec.~\ref{sec:exper} offers an overview of the employed experimental techniques, including the relevant settings and conditions. 
In Sec.~\ref{ssec:nmr} the electronic correlations in the normal phase 
are studied via NMR experiments, whereas 
in Sec.~\ref{ssec:muSR} the SC phase is investigated by $\mu$SR and the respective main SC
 parameters are evaluated quantitatively. Finally, Sec.~\ref{sec:conclusions} offers conclusions and a
 general discussion regarding open questions about the class of weakly-correlated NCSs, with 
particular reference to SrAuSi$_3$.

\vspace{-5pt}\section{Experimental Details}\label{sec:exper}\enlargethispage{8pt}

\begin{figure}[htb]
\includegraphics[width=0.9\linewidth]{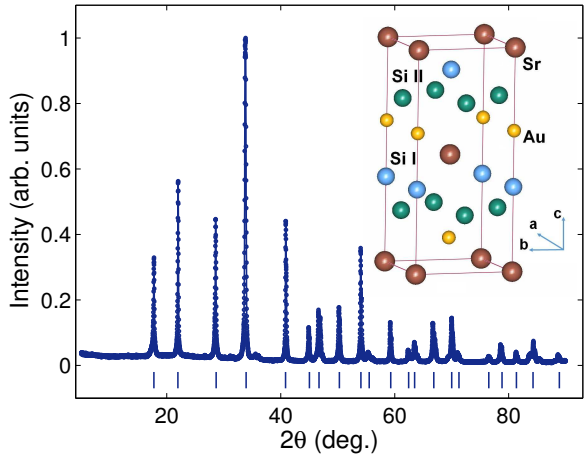}%
\caption{\label{fig:xray}XRD pattern of the polycrystalline sample. Data 
were acquired using the $K_\alpha$ line of Cu (40 kV, 15 mA) and fitted 
by following a standard Rietveld-refinement procedure.}
\end{figure}

Samples of SrAuSi$_3$ were prepared using the
solid-state reaction method under 6\,GPa at 1000--1200$^\circ$C.\cite{Isobe2014, Isobe2016}
Preliminary X-ray diffraction (XRD) measurements were used to verify the 
sample's crystal structure (see Fig.~\ref{fig:xray}). No spurious phases above a 0.1\% limit could be 
detected and the already known tetragonal $I4mm$ structure was confirmed.\cite{Isobe2014}

The NMR investigations included line-shape and spin-lattice relaxation ($T_1$) measurements 
in an applied magnetic field of 7.0571\,T. The most suitable nucleus for our study was $^{29}$Si 
(spin $I=1/2$,  natural abundance 4.7\%, and a Larmor frequency of 59.685\,MHz).
The $^{29}$Si NMR spectra were obtained via the 
fast Fourier transformation (FFT) of the spin-echo signals generated by 
$\pi/2$--$\pi$ rf pulses of 11 and 22-$\mu$s duration, respectively, with 
a 50-$\mu$s spacing.
The recycle delays ranged from 0.2\,s at room temperature up to 10\,s at 2\,K
To improve the quality of the spectra, frequency sweeps with 20-kHz steps were 
used. The nuclear spin-lattice relaxation times $T_1$ were measured at both peak frequencies, corresponding to
two inequivalent Si-sites, by using the inversion-recovery method with spin-echo detection at variable delays. 
 Due to the intermediate sensitivity of $^{29}$Si nuclear spins, several scans had to be accumulated 
 (from 200 at 2\,K up to $1\times10^4$ at 290\,K).

\begin{figure}[htb]
\includegraphics[width=0.9\linewidth]{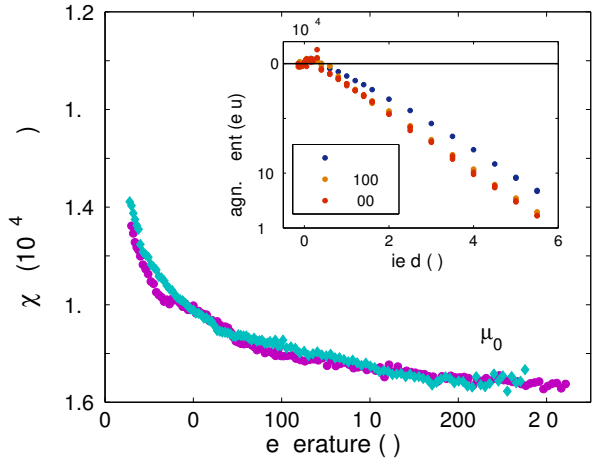}%
\caption{\label{fig:SQUID} Magnetometry $\chi_m(T)$ data measured at $\mu_0H$ = 5\,T in the normal phase (down to 4\,K). The inset shows the magnetization $M(H)$ curves at three selected temperatures (5, 100, and 300\,K).}
\end{figure}

The local-field distribution within the material was monitored with the very sensitive $\mu$SR technique. The transverse-field (TF) $\mu$SR technique is often used to measure the value and the temperature dependence of the magnetic-penetration depth $\lambda$ in the vortex state of type-II superconductors~\cite{Sonier2000}. $\lambda^{-2}(T)$ is proportional to the density of superconducting carriers $n_{\rm s}$, hence it provides information on the character of the gap in the electronic excitation spectrum below $T_c$. The TF-$\mu$SR experiments were carried out at the low-temperature (LTF) instrument of the $\pi$M3 beam line at the Paul Scherrer Institute (Villigen, Switzerland). To avoid flux-pinning effects, the sample was cooled to the base temperature in an applied magnetic field of 20\,mT. The data were collected between 0.02 and 1.8 K while warming the sample in a steady field. Typically $\sim12 \times 10^{6}$ 
muon-decay events were collected for each spectrum. The TF-$\mu$SR data were analyzed by means of the software package MUSRFIT~\cite{Suter2012}.

Given the low $T_{c} = 1.54$\,K and the modest upper critical field of 
0.18\,T, an NMR study of SrAuSi$_3$ in the superconducting 
phase is quite challenging. The required spin-polarizing field suppresses 
superconductivity in fields exceeding $H_{c2}(0)$. If the chosen applied field were 
substantially weaker, the ${}^{29}$Si Larmor frequency would have been very low, 
resulting in an extremely poor S/N ratio. On the other hand, due to the intrinsic spin-polarization 
of muon beams, TF-$\mu$SR 
can be successfully used in the SC phase even at small applied 
fields (ca.\ 20\,mT), yet it is much less sensitive in the normal phase.
In view of these considerations, we chose NMR to probe 
the normal phase and $\mu$SR for the SC-phase of SrAuSi$_3$.

\vspace{-5pt}\section{Results and Discussion}\label{sec:results}\enlargethispage{8pt}

\subsection{\label{ssec:nmr}Normal-phase nuclear magnetic resonance investigations}

For a preliminary characterization of the electronic and magnetic properties of the 
normal phase, we measured the sample's magnetization
$M(T,H)$ with a commercial MPMS (magnetic property measurement system) XL setup.
With a typical $\chi_\mathrm{dia}$ of $-1.6 \times 10^{-4}$ \,cm$^3$/mol 
(see Fig.~\ref{fig:SQUID}), the $\chi_m(T) = M(T)/H$ data of SrAuSi$_3$ mimic 
those of SrPt$_3$P,\cite{Shiroka2015} a centrosymmetric superconductor and of 
the diamagnetic Au ($-1.2 \times 10^{-4}$ \,cm$^3$/mol).\cite{Mendelsohn1970}
The inset of Fig.~\ref{fig:SQUID} confirms that the bulk of the sample is predominantly 
diamagnetic. The small positive response at low fields suggests the presence of 
a small amount ($<0.1$\%) of impurities, reflected also in the low-field hysteretic 
cycle (not shown). 

\begin{figure}[htb]
\includegraphics[width=0.9\linewidth]{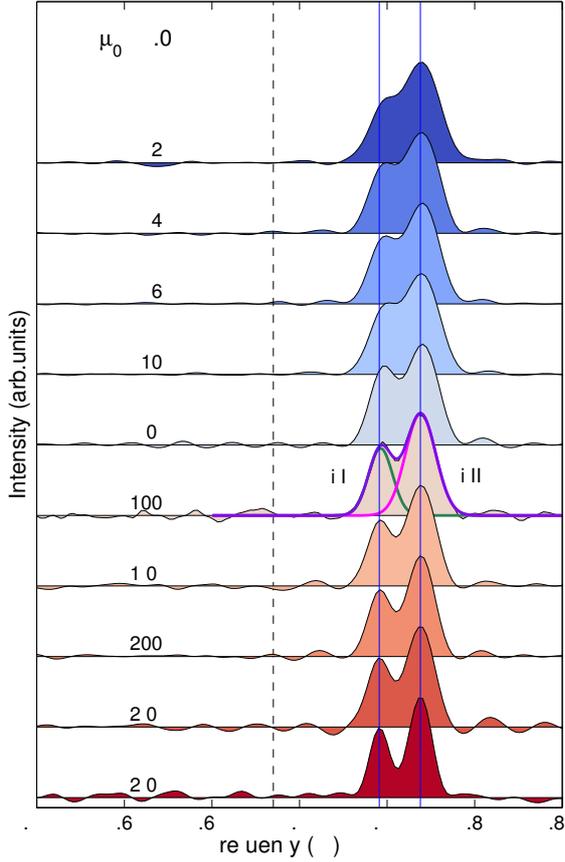}%
\caption{\label{fig:lines}$^{29}$Si line-shapes measured at 7.06\,T. The spectra 
were fitted by a two-Gaussian model, as shown for $T = 100$\,K. The dashed black line corresponds to the 
reference Larmor frequency of $^{29}$Si, whereas the solid blue lines correspond 
to the frequencies of the two peaks at room temperature (RT).}
\end{figure}

From the published structure-data file,\cite{Isobe2014} Si occupies two 
inequivalent lattice sites: Si-I (two atoms per unit cell, i.e., 
one internal and four at the edges) and Si-II (four atoms per unit cell, i.e.,  
eight atoms on the faces). From the 2:4 multiplicity ratio, we 
expect two resonance lines, related to the Si-I and Si-II nuclei, with 
an ideal ratio of 2 for the peak areas.
The observed spectra were fitted by assuming a simple model (see 
Eq.~\ref{eq:gauss_eq}), consisting of the sum of two Gaussian peaks, 
as illustrated in Fig.~\ref{fig:lines} for $T = 100$\,K:
\begin{equation}
I(f) = a_1 \cdot \exp\left[ - \frac{(f-f_1)^2}{2\sigma_1^2}\right] 
     + a_2 \cdot \exp\left[ - \frac{(f-f_2)^2}{2\sigma_2^2}\right],
\label{eq:gauss_eq}
\end{equation}
where $a_i$ ($i = 1,2$) are the normalization factors, $f_i$ the peak
maxima, and $w_i$ their full widths at half maximum (FWHM), with 
$w_i = 2(2 \ln 2)^{1/2}\sigma_i \simeq 2.35\,\sigma_i$.

\begin{figure}[htb]
\includegraphics[width=0.9\linewidth]{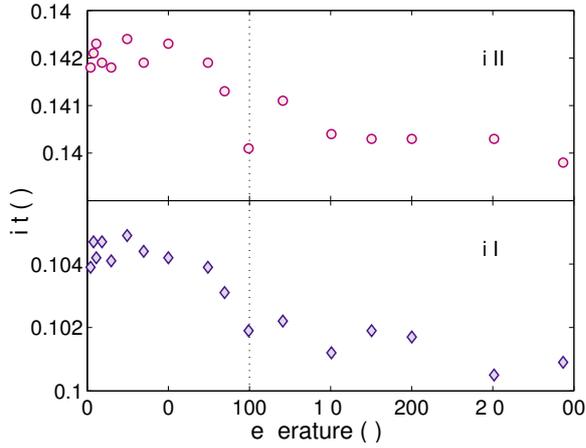}
\caption{\label{fig:shift}Shifts of the two $^{29}$Si NMR lines vs.\ temperature.}
\end{figure}

As the temperature is lowered from 300 to 5\,K, the shift of 
the Si-I NMR line varies from 0.1 to 0.105\%, whereas that of the 
Si-II lines varies from 0.14 to 0.142\% (see Fig.~\ref{fig:shift}).
These shift values are similar to those reported in literature for 
typical metals such as Na.\cite{Narath1968} 
 Indeed, a non-varying value in the normal phase is commonly observed
in simple metals whose electronic susceptibility $\chi_P$ varies only weakly with temperature.\cite{Shiroka2015}
In both cases, a tiny gradual enhancement of the line shift is observed below 100\,K, 
probably due to a small change in the electronic environment (see Fig.~\ref{fig:shift}).

\begin{figure}[htb]
\includegraphics[width=0.9\linewidth]{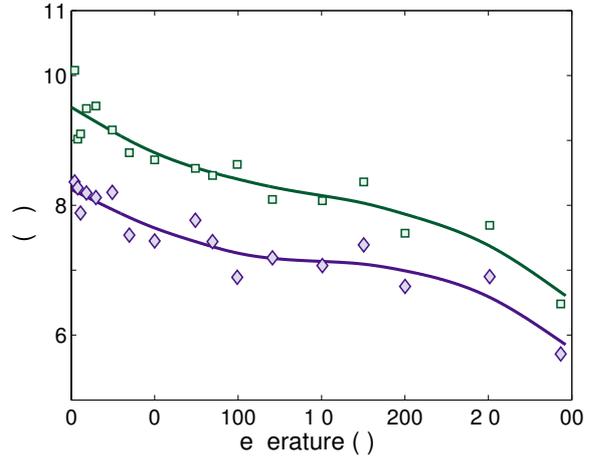}
\caption{\label{fig:fwhm}FWHM vs.\ $T$ plots of both  $^{29}$Si NMR lines 
indicate a gradual broadening as the temperature is lowered.}
\end{figure}

As shown in Fig.~\ref{fig:fwhm}, both NMR lines broaden upon 
lowering the temperature, with FWHM values changing smoothly from about 
6 to 8\,kHz for Si-I and from 7 to 9\,kHz for Si-II sites.
From the fits we estimate a ratio of peak areas of $1.8\pm0.2$, to be compared
with the expected value of 2, as explained above. 
The small discrepancy is most likely due to a $T_1$-induced 
signal-intensity modulation, reflecting the different relaxation rates of 
nuclei at the Si-I and Si-II sites, whose $T_1$ values differ by a factor of 2 at 
room temperature. 

\begin{figure}[htb]
\includegraphics[width=0.9\linewidth]{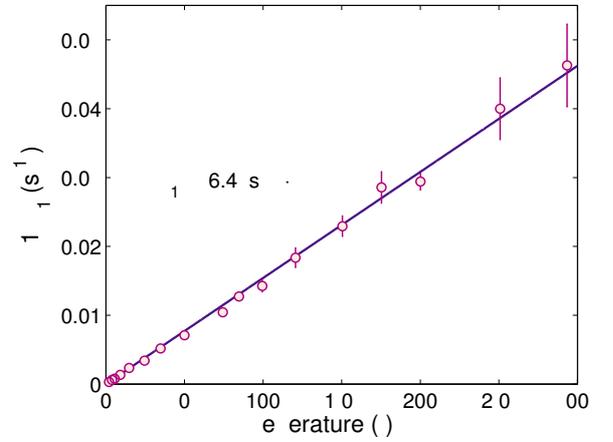}
\caption{\label{fig:invT1T}$1/T_1$ vs.\ $T$ for the $^{29}$Si NMR of Si-II site.}
\end{figure}
In the normal phase we find a linear variation of $1/T_1(T)$, i.e., a Korringa 
behavior typical of simple metals (see Fig.~\ref{fig:invT1T}). The average 
$T_1T$ value is 6.49\,sK which, by considering the measured 0.14\% Knight shift, 
implies an experimental Korringa constant $S_{\mathrm{exp}} \equiv T_1TK^2 = 1.31\times 10^{-5}$\,sK, 
twice the theoretical value 
$S_0 = \hbar(\gamma_e/\gamma_{\mathrm{Si}})^2 /(4\pi k_{\mathrm{B}}) = 6.66 \times 10^{-6}$\,sK. 
From the relative Korringa ratio $\mathcal{K} \equiv S_0/S_{\mathrm{exp}} = 0.508$, one can evaluate the enhancement of the static and dynamic spin susceptibilities, driven by the $q$-independent electron-electron interaction potential\cite{Pines1954} $\zeta = \hbar^2 {\gamma_e}^2 \alpha/ 2\pi \chi_{s0}$, where $\gamma_e$ is the gyromagnetic ratio of the electron, $\chi_{s0}$ the spin susceptibility, and $\alpha$ is a parameter that is uniquely defined by the value of $\mathcal{K}$ via:\cite{Narath1968,Walstedt2008} 
\begin{equation}
\mathcal{K}(\alpha) = 2(1-\alpha)^2 \int_{0}^{1} x/[1-\alpha G(x)]^2 \mathrm{d}x.
\label{eq:Korringarelratioeq}
\end{equation}
The quantity $G(x)$ in Eq.~(\ref{eq:Korringarelratioeq}) is the Lindhard function for free electrons, i.e., $G(x) = \frac{1}{2}\{1+[(1-x^2)/2x]\ln(|1+x|/|1-x|)\}$, with the normalized momentum $x = q/(2k_\mathrm{F})$, where $k_\mathrm{F}$ is the Fermi momentum under the assumption of a spherical Fermi surface.
An analytical evaluation of the $\zeta$ parameter is not possible, since the value of the spin susceptibility $\chi_{s0}$, i.e. the Pauli susceptibility, generally cannot be estimated from the measured magnetic susceptibility $\chi_m = M(T,H)/H$, which contains also additional contributions due to, e.g., the core electrons and the electronic correlations.
Nevertheless, the experimentally determined $\mathcal{K}(\alpha)$ suggests the presence of ferromagnetic correlations in SrAuSi$_3$ which, however, are incompatible with the rather weak Knight shift. From the experimental $\mathcal{K}(\alpha)$ value, one can solve Eq.~(\ref{eq:Korringarelratioeq}) via numerical integration and obtain $\alpha = 0.34(3)$. Such an $\alpha$ value indicates weak but not negligible electron-electron interactions, in line with the Moriya theory 
of exchange enhancement.\cite{Moriya1963}
Indeed, similar $\alpha$ and  $\mathcal{K}(\alpha)$ values are found in many alkali metals, 
as well as in Cu and Ag, all well-known realizations of the nearly-free electron model.\cite{Narath1968,Walstedt2008}

\begin{figure}[htb]
\includegraphics[width=0.8\linewidth]{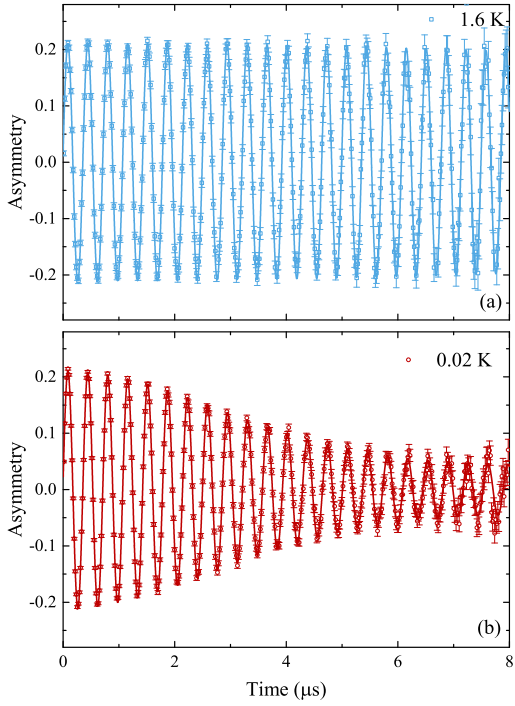}
\caption{(a) and (b) show the TF-$\mu$SR time spectra of SrAuSi$_3$, collected at 1.6 and 0.02 K (above and below $T_{\rm c}$) in an applied field of 20 mT. The solid lines are least-square fits to the asymmetry spectra according to Eq.~(\ref{eq:TFequation}).}
 \label{fig:Asy_TF}
\end{figure}

\subsection{\label{ssec:muSR}Muon-spin rotation in the superconducting phase}
TF-$\mu$SR measurements were carried out to investigate the superconducting properties of SrAuSi$_3$. Figure~\ref{fig:Asy_TF}(a) and (b) show the TF-$\mu$SR time spectra, collected at 1.6 and 0.02\,K in a magnetic field of 20~mT. While the $\mu$SR time spectrum obtained at 1.6\,K in the normal state show almost no relaxation, data collected at 0.02 K in the superconducting state reveal pronounced damping in the $\mu$SR time spectra due to the inhomogeneous field distribution generated by the formation of a vortex lattice. The TF-$\mu$SR time spectra were analyzed using the oscillatory Gaussian decay function:\cite{Sonier2000}
\begin{multline}
A^{\mathrm{TF}}(t)=A(0)\exp\left(-\sigma^{2}t^{2}\right/2)\cos\left(\gamma_\mu B_{\rm int} t +\phi\right) \\
+A_{\rm bg}(0)\cos\left(\gamma_\mu B_{\rm bg}t +\phi\right),
\label{eq:TFequation}
\end{multline}
where the fit parameters $A(0)$ and $A_{\rm bg}$(0) are the initial asymmetries of the sample and background signals, $\gamma_{\mu}/2\pi=13.55$~kHz/G is the muon gyromagnetic ratio~\cite{Sonier2000}, $B_{\rm int}$ and $B_{\rm bg}$ are the internal and background magnetic fields, $\phi$ is the initial phase of the muon precession signal, and $\sigma$ is the Gaussian muon-spin relaxation rate. The last term in Eq.~(\ref{eq:TFequation}) represents the background signal that originates mostly from muons hitting the silver sample holder and is considered as non-relaxing within the muon time window.\cite{Sonier2000} The signal to background ratio, $A_{\rm bg}(0)/A(0)=0.21$.

\begin{figure}[htb]
\includegraphics[width=0.8\linewidth]{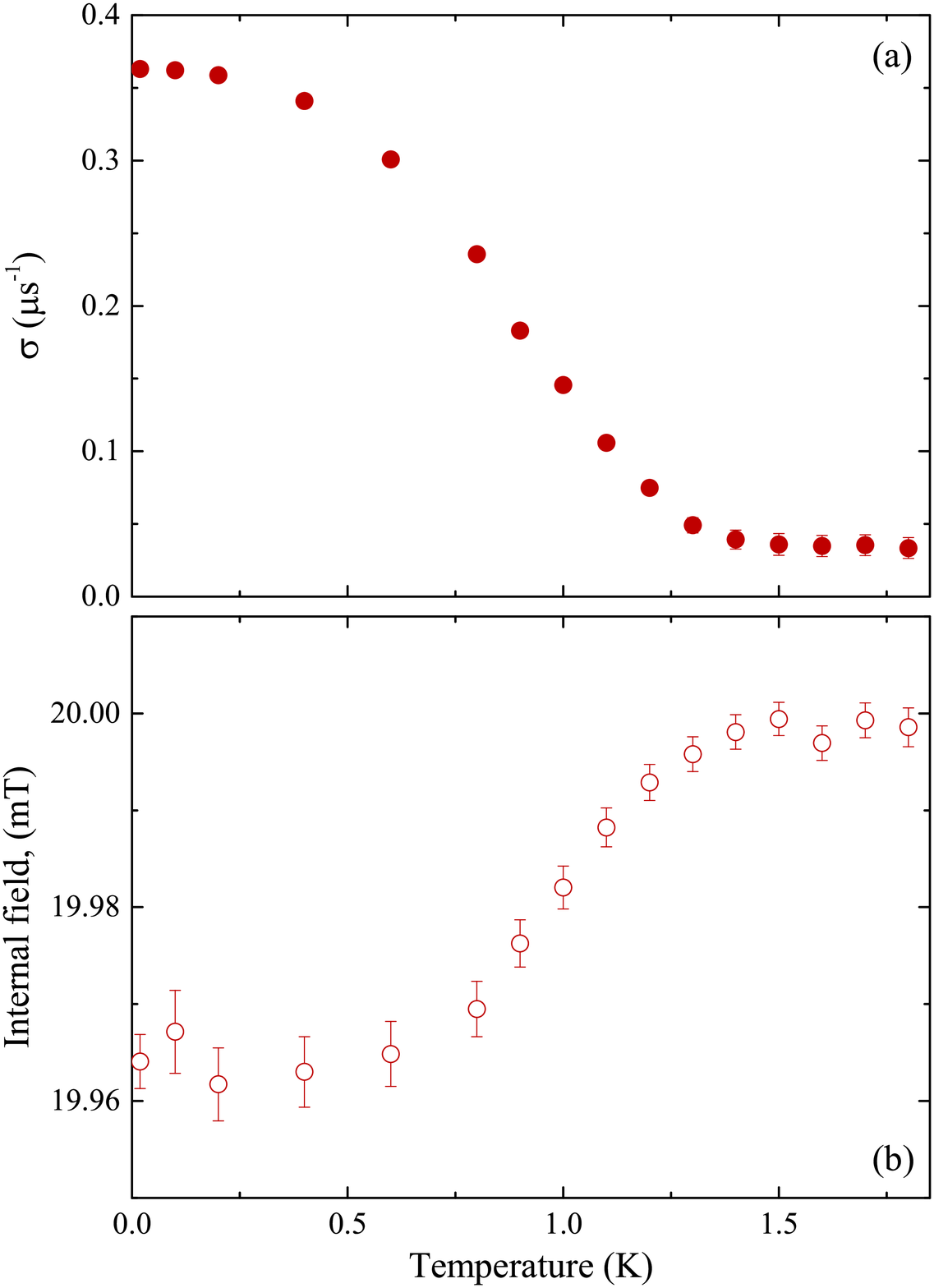}
\caption{(a) Temperature dependence of the muon depolarization rate $\sigma$ of SrAuSi$_3$ in an applied magnetic field of 20 mT. (b) Diamagnetic shift of the internal field experienced by the muons just below $T_{\rm c}$.}
 \label{fig:sigmaT}
\end{figure}

Figure~\ref{fig:sigmaT}(a) shows the temperature dependence of $\sigma$ of SrAuSi$_3$ for an applied field of 20\,mT, exhibiting a pronounced increase just below $T = T_{\rm c}$. Figure~\ref{fig:sigmaT}(b) shows the temperature dependence of the internal magnetic field at the muon site, revealing the expected diamagnetic shift and confirming the bulk-type superconductivity in this material. The superconducting contribution $\sigma_\mathrm{sc}$ can be obtained by quadratically subtracting the nuclear moment contribution $\sigma_{\rm nm}$ (measured above $T_{\rm c}$) from the total $\sigma$, as ${\sigma_{\rm sc}}^2=\sigma^2-{\sigma_{\rm nm}}^2$. Since we do not expect any structural transition in SrAuSi$_3$ over the temperature range of the TF-$\mu$SR study, $\sigma_{\rm nm}$ is assumed to be temperature independent. In an isotropic type-II superconductor with a hexagonal Abrikosov vortex lattice, the magnetic penetration depth $\lambda$ is related to $\sigma_{\mathrm{sc}}$ by the Brandt equation~\cite{Brandt2003}:
\begin{multline}
\sigma_{\mathrm{sc}}[\mu{\rm s}^{-1}](T)=4.854\times10^4\left(1-\frac{H}{H_{\rm c2}(T)}\right) \\
\left[1+1.21\left(1-\sqrt{\frac{H}{H_{\rm c2}(T)}}\right)^3\right]\lambda^{-2}[{\rm nm}^{-2}],
\label{eq:Brandt_equation}
\end{multline}
where $H$ and $H_{\rm c2}(T)$ are the applied and upper critical fields, respectively. 

\begin{figure}[htb]
\includegraphics[width=0.9\linewidth]{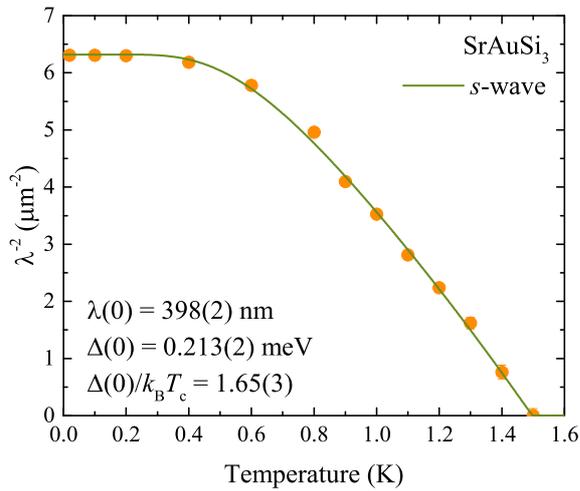}
\caption{The temperature dependence of $\lambda^{-2}(T)$. The solid line is a fit to the $\lambda^{-2}(T)$ with the weak-coupling BCS $s$-wave model.}
 \label{fig:lambdaT}
\end{figure}

Figure~\ref{fig:lambdaT} shows $\lambda^{-2}(T)$ of SrAuSi$_3$, calculated using Eq.~(\ref{eq:Brandt_equation}). The temperature dependence of $H_{\rm c2}(T)$ was taken from Ref.~\onlinecite{Isobe2014}. $\lambda^{-2}$ is proportional to the effective superfluid density, $\rho_{\rm s}$. 
Therefore, $\lambda^{-2}(T)$ provides information about the magnitude of the superconducting gap and the symmetry of the gap structure. Figure~\ref{fig:lambdaT} reveals that $\rho_{\rm s}$ of SrAuSi$_3$ is nearly constant below $T_{\rm c}/3\approx0.5$ K, suggesting a nodeless superconducting gap in this material. A good fit to $\lambda^{-2}(T)$ can be achieved with a single-gap BCS $s$-wave model~\cite{Tinkham1975,Prozorov2006} (solid line in Fig.~\ref{fig:lambdaT}) using the functional form:

\begin{equation}
\frac{\lambda^{-2}(T)}{\lambda^{-2}(0)}= 1+2\int_{\Delta(T)}^{\infty}\left(\frac{\partial f}{\partial E}\right)\frac{E \mathrm{d} E}{\sqrt{E^2-\Delta(T)^2}}.
 \label{eq:lambda}
\end{equation}

Here $\lambda^{-2}(0)$ is the  zero-temperature value of the magnetic penetration depth and $f=[1+\exp(E/k_{\mathrm{B}}T)]^{-1}$ denotes the Fermi function. The BCS temperature dependence of the superconducting gap function is approximated by~\cite{Carrington2003}
\begin{equation}
\Delta(T)=\Delta(0)\tanh\left\{1.82\left[1.018\left(\frac{T_c}{T}-1\right)\right]^{0.51}\right\},
 \label{eq:delta}
\end{equation}
where $\Delta(0)$ is the gap magnitude at zero temperature. The fit yields $T_c=1.49(1)$~K, $\lambda(0)=398(2)$~nm, and $\Delta(0)=0.213(2)$~meV. The ratio $\Delta(0)/k_{\rm B}T_{\rm c}=1.65(3)$, which is lower than the BCS value of 1.76, confirms that SrAuSi$_3$ is a weak-coupling superconductor.

\section{Conclusions}\label{sec:conclusions}

A combined study of the normal and superconducting phase of SrAuSi$_3$ was performed through NMR and $\mu$SR in TF mode measurements, respectively. In the normal phase this compound exhibits a Korringa behavior with enhanced electron-exchange behavior, in line with the Moriya 
theory mentioned above. The temperature dependence of the London penetration depth $\lambda$ was evaluated from the TF-$\mu$SR time spectra. $\lambda (T)$ can be modelled well by assuming a single-gap BCS $s$-wave scenario with $\Delta(0)=0.213(2)$~meV which suggests that SrAuSi$_3$ is a nodeless superconductor. The gap-to-$T_c$ ratio provides evidence for weak-coupling superconductivity in SrAuSi$_3$. The magnetic penetration depth was estimated as $\lambda(0)=398(2)$~nm.

Single-band fully-gapped BCS superconductivity and weak electronic correlation confirm the small splitting of the bands in the vicinity of the Fermi level $E_\mathrm{F}$, visible only in proximity of the Brillouin zone edges.\cite{Arslan2016} This hints at a small ASOC that, along with weak correlations, may hinder the singlet-triplet mixing.
Although a mixed-parity pairing is predicted to be a general feature of 
NCSs with ASOC, many weakly correlated compounds exhibit a rather simple 
behavior,\cite{Smidman2017} as reported here for the SrAuSi$_3$ case.
This general trend can be explained by considering that the triplet component is very weak, therefore, two gaps of very similar magnitude will be dominated by spin-singlet pairing, i.e., a mechanism indistinguishable from ordinary $s$-wave superconductors. 
When looking for new systems exhibiting singlet-triplet mixing, a 
simple comparison of the relative energies of ASOC and SC condensate, 
is often not sufficient, since also pairing interactions in both the singlet 
and triplet channels are required for the occurrence of parity mixing.
This is evidenced by the fact that weakly correlated NCSs do not exhibit mixing, 
whereas their strongly correlated counterparts (including Ce- and other 
heavy-electron based systems) may do so.\cite{Smidman2017}

In perspective, further experimental and theoretical work is required 
in order to confirm the occurrence of mixed singlet-triplet pairing, 
since the role of ASOC and that of the strength of electronic correlations 
still remains poorly understood. Furthermore, due to their non-trivial 
band structure, topological SC states are expected in several NCSs. 
Investigations by local probes, as the ones reported here, 
remain crucial to support or rule out new hypotheses 
aiming to explain the intriguing properties of noncentrosymmetric 
superconductors.

\vspace{-5pt}
\begin{acknowledgments}
This work was financially supported in part by the 
Schwei\-ze\-rische Na\-ti\-o\-nal\-fonds zur F\"{o}r\-de\-rung 
der Wis\-sen\-schaft\-lich\-en For\-schung (SNF) and by the Japan 
Society for the Promotion of Science (JSPS) through Grants-in-Aid 
for Scientific Research (Grant No.\ 16K06712).
\end{acknowledgments}


%

\end{document}